\documentclass[12pt,preprint]{aastex}

\newcommand{\cmii}{\mbox{cm$^{-2}$}}
\newcommand{\cmiii}{\mbox{cm$^{-3}$}}
\newcommand{\cms}{\mbox{cm~s$^{-1}$}}
\newcommand{\cog}{CO~G138}
\newcommand{\cogfull}{CO~G138.1$-$1.9$-$48}
\newcommand{\coi}{$^{12}$CO($J=1-0$)}
\newcommand{\ergs}{\mbox{erg~s$^{-1}$}}
\newcommand{\g}{\mbox{g}}
\newcommand{\gcms}{\mbox{g~cm~s$^{-1}$}}
\newcommand{\gcmss}{\mbox{g~cm~s$^{-2}$}}
\newcommand{\hig}{HI~G137}
\newcommand{\higfull}{HI~G137.7$-$1.6$-$60}
\newcommand{\gs}{\mbox{g~s$^{-1}$}}
\newcommand{\ha}{\mbox{\sc{H}{$\alpha$}}}
\newcommand{\hi}{\mbox{H\,{\sc i}}}
\newcommand{\hii}{\mbox{H\,{\sc ii}}}

\newcommand{\kelvin}{\mbox{K}}
\newcommand{\kms}{\mbox{km~s$^{-1}$}}
\newcommand{\kpc}{\mbox{kpc}}
\newcommand{\mdot}{\mbox{$\dot{M}$}}
\newcommand{\mhz}{\mbox{MHz}}
\newcommand{\msun}{\mbox{M$_{\sun}$}}

\newcommand{\msunyr}{\mbox{M$_{\sun}$~yr$^{-1}$}}
\newcommand{\pc}{\mbox{pc}}
\newcommand{\pckms}{\mbox{pc~(km~s$^{-1}$)$^{-1}$}}
\newcommand{\pers}{\mbox{s$^{-1}$}}
\newcommand{\psf}{\mbox{s$^{-1}$~cm$^{-2}$~ster$^{-1}$}}
\newcommand{\yr}{\mbox{yr}}

\slugcomment{To appear in {\it The Astrophysical Journal}}

\shorttitle{Unusual {\hi} Filaments}
\shortauthors{Knee, Wallace, and Normandeau}

\begin{document}

\title{An Unusual System of {\hi} Filaments near WR~5 and HD~17603}

\author{Lewis B.G. Knee}
\affil{Millimetre Astronomy Group, Herzberg Institute of
       Astrophysics, \\ National Research Council of Canada, \\
       5071 West Saanich Road, Victoria, BC, V9E 2E7, Canada}
\email{lewis.knee@nrc-cnrc.gc.ca}

\author{Bradley J. Wallace}
\affil{Defence Research and Development Canada, \\
       3701 Carling Avenue, Ottawa, ON, K1A 0Z4, Canada}
\email{brad.wallace@drdc-rddc.gc.ca}

\and

\author{Magdalen Normandeau}
\affil{Department of Physics, Amherst College, \\
       Amherst, MA, 01002-5000, United States of America}
\email{mnormandeau@amherst.edu}

\begin{abstract}

We report the discovery of a system of unusual {\hi} filaments which appear to
be associated with molecular clouds in the Perseus spiral arm of our Galaxy. We
investigate the hypothesis that this system is the result of a directed flow of
dissociated gas from clouds trapped within an extended wind flow from massive
stars. The Wolf-Rayet star WR~5 and the OIb(f) star HD~17603 are identified as
candidate driving sources. However, an examination of this hypothesis within
the context of the theory of mass-loaded winds shows 
that these two stars alone can not account for the energetics and kinematics of the
required spherically symmetric wind flow. Unless the apparent association
between {\hi}, molecular gas, and stars is an accidental one, we suggest that
other as-yet unidentified stars must have contributed to driving the filaments.

\end{abstract}


\section{Introduction}\label{sect:intro}

Massive O-type stars, including the progenitors of Wolf-Rayet stars, inject
enormous quantities of energy into the interstellar medium (ISM) in the form of
dissociating and ionizing radiation and winds. The radiation fields create
circumstellar/interstellar {\hi} and {\hii} regions into which winds blow,
creating large-scale swept-up expanding shells of gas known as stellar wind
bubbles \citep{dys72,koo92a,koo92b}. There are numerous analytical and
numerical studies of the formation, structure, and evolution of stellar wind
bubbles (SWBs), but there are few convincing observations of {\hi} bubbles
around Wolf-Rayet stars and massive O-stars. The VIIth catalogue of galactic
Wolf-Rayet stars indicates that only $\approx 15\%$ are associated with {\hi}
bubbles \citep{van01}. Surveys are incomplete, but many of these proposed
bubbles are not securely established. 

The interstellar environment of massive stars is observed to be extremely
complex and is likely to be the combined result of inhomogeneous initial
conditions and the influence of the stars. Our understanding of the physics of
astrophysical flows in inhomogeneous media is incomplete \citep{har93}. In
particular the effects of conduction, diffusion, turbulence, magnetism, and of
mixing and mass-loading processes are all poorly understood. Therefore, in the
observational study of SWBs and the environment of massive stars it is
important to remain open to the possibility of phenomena not predicted by
standard SWB models.

In this paper we present one possible example of such a phenomenon. We
investigate the nature of a large-scale system of neutral atomic hydrogen
filaments which apparently emanate from compact molecular clouds located in or
near the
Perseus arm of our Galaxy. We investigate the possibility that these {\hi}
``tails'' are the result of molecular cloud mass-loading within the wind flow
of massive stars. 

The observations and data processing are briefly outlined in the next section.
Section~\ref{sect:results} describes the observational results. In
section~\ref{sec:driving_srcs} we consider possible driving sources, and in
section~\ref{sec:environ} we take a closer look at the environment of the two
main candidates. The discussion follows (\S\ref{sec:discuss}), in which we
first consider the kinematics and timescales involved (\S\ref{sec:kin_time}),
and then present an analysis within the framework of a mass-loaded flow model
(\S\ref{sec:mass-loaded}). Our conclusions follow in
section~\ref{sec:conclude}.

\section{Observations and Data Processing}\label{sect:obs}

The data discussed in this paper were obtained as part of the Canadian Galactic
Plane Survey, a long-term multi-wavelength imaging survey of
the ISM of the
northern Galactic plane \citep{tay03}. Table~\ref{tbl:CGPS_param} summarizes
the main parameters of the relevant CGPS data. 
The principal observational components of the CGPS are arcminute resolution
mosaics of 21~cm (in all Stokes parameters) and 74~cm (right hand circular
polarization) radio continuum, and 21~cm {\hi} line emission, all observed with
the Synthesis Telescope \citep{lan00} at the Dominion Radio Astrophysical
Observatory of the Herzberg Institute of Astrophysics.

Complementary data produced as part of the CGPS include mosaics of dust
emission created using HIRES-processed IRAS data \citep{cao97,ker00} and of
{\coi} emission from the FCRAO Outer Galaxy Survey \citep{hey98}. The
CGPS CO mosaics are the result of a reprocessing of the original OGS data,
wherein correlated noise between pixels caused by the observing method was
suppressed, and faint contaminating emission found in the reference positions
of the survey was removed \citep{bru00}. The reprocessed data were resampled
and projected into the CGPS mosaic system.

\section{Results}\label{sect:results}

\subsection{Neutral Atomic Hydrogen}

A mean brightness temperature {\hi} image, averaged between LSR radial
velocities $-53.4$ and $-66.6$~{\kms}, is shown in Figure~\ref{fig:mean_Tb_HI}. 
Prominent in this image is a linear {\hi} structure $1\fdg8$ long, running at
an angle $\sim 20\degr$ north of west relative to the Galactic plane. In the
extreme southeast this structure is a relatively narrow ($\sim 2\arcmin$ wide)
filament beginning near $(l, b)=(138\fdg7, -2\fdg0)$. Near $(138\fdg2,
-1\fdg9)$, the narrow filament is joined by a broader ($\sim 0\fdg5$) structure
which runs parallel and to the south of it. This structure continues to broaden
towards the northwest, reaching a maximum width of $\sim 0\fdg8$. Its northwest
extremity lies near $(136\fdg7, -1\fdg0)$, where it blends into larger scale
{\hi} structures in the Galactic plane. The approximate centre of the linear
structure is at $(l, b) \approx (137\fdg7, -1\fdg6)$; we designate it
{\higfull}, and for brevity refer to it as {\hig} in this paper. 

The radial velocity width of {\hig} is $15$~{\kms}, and there is a distinct
velocity gradient along its length: velocities become more negative towards the
northwest. This velocity gradient clearly separates {\hig} from the Galactic
{\hi} emission to the northwest (Figure~\ref{fig:vel_gradient}). Individual 
channel maps (Figure~\ref{fig:channel_maps}) hint at the existence of linear
substructures, resulting in the overall impression that {\hig} is composed of
several parallel filaments. We can detect no clear velocity structure
in the direction transverse to the length of the filaments: this may be due to
a combination of sensitivity and angular resolution limitations for 
individual filaments and confusion between them.

We are unable to make a direct measurement of the distance to {\hig}. There
are no
known {\hii} regions in the direction of the filaments, and there are
large discrepancies between kinematic and photometric distances in the Galactic
plane in this region in the sense that kinematic distances can be 
greatly over-estimated compared to the latter \citep{fos03}.

\subsection{Carbon Monoxide}

The CGPS CO data were searched for molecular clouds having positional and/or
velocity coincidences with {\hig}. We found no CO emission near {\hig} within
the velocity range of the {\hi} structure. Widening the velocity range of our
search revealed an irregular chain of CO clouds in the southeast of {\hig} at
$-44 \la v \la -50$~{\kms} extending $\sim 1\fdg4$ in a direction perpendicular
to the length of the {\hi} structure. Of these, the cloud with the brightest
peak CO line brightness (${{\rm T}_{\rm R}}^{*} = 4.0$~K), which we dub
{\cogfull} ({\cog} for brevity), lies at the convergence point of the
broad-structured component of {\hig} (Figure~\ref{fig:co_clouds}). The velocity
difference between {\cog} and {\hig} ranges between 5 and 20~{\kms}. The
northern narrow filament of {\hig} appears to emerge from between a gap in the
chain of CO clouds. Within this gap there is a faint (1.3~K) compact cloudlet
at $-47$~{\kms} embedded within very faint ($\sim 0.5$~K) diffuse CO emission.

\section{Search for Driving Sources}\label{sec:driving_srcs}

The positional association between molecular clouds and the convergent point of
the {\hi} filaments suggests that {\hig} might be the result of a directed flow
of dissociated gas from the molecular clouds. If {\hig} is associated with the
CO clouds at $\sim -48$~{\kms}, its distance is $D \sim 2$~kpc, the approximate
distance to the Perseus Arm at this longitude. The source(s) of the radiation
and/or winds driving {\hig} will be at the same distance. The former would have
to be located on the sky such that the {\hi} filaments point towards them and
the CO clumps lie between them and the filaments in a more or less collinear
fashion. The SIMBAD database was used to search for O, B, and Wolf-Rayet type
stars within $0\fdg5$ of $(l, b) = (138\fdg5, -2\fdg0)$
(Table~\ref{tbl:stars}). Of the resulting ten stars, the Wolf-Rayet star WR~5
and the O-type supergiant HD~17603 currently have substantial stellar winds and
ionizing radiation. These two therefore merit closest attention.  

\subsection{Wolf-Rayet~5 and HD~17603}

WR~5 (HD~17638) is a Wolf-Rayet star of spectral type WC6 \citep{van01}. 
It has a reddening of ${\rm E}_{{\rm B}-{\rm V}} = 0.96$ (${\rm A}_{\rm V} \sim
3$) and a photometrically derived distance of 1.91~kpc \citep{van01}. 

HD~17603 is an O7.5Ib(f) star \citep{HD17603type} located $7\farcm5$ northwest
of WR~5. Stars of type O(f) are thought to be precursors of Wolf-Rayet stars. 
The available UBV photometry for this star \citep{hau70} is consistent with a
reddening ${\rm E}_{{\rm B}-{\rm V}} = 0.93$ (${\rm A}_{\rm V} \approx 3$). The
absolute magnitudes of Of stars are not well determined: however assuming that
this star has ${\rm M}_{\rm V} \approx -6.3$ \citep{gar82}, it lies at a
distance $D \sim 2.2$~kpc. 

Both WR~5 and HD~17603 satisfy the positional criteria to be considered
candidate driving sources. In addition, both are hot, luminous objects with
strong winds. For both stars, the apparent alignment of
{\hi}$-$CO$-$star is very
good, and these stars are likely to be the most luminous and energetic of those we
have identified. Finally, their approximate distances are consistent with that
proposed for {\hig}. Stellar wind parameters for both stars are
listed in Table~\ref{tbl:WR5_HD17603}.

A final consistency check on the stellar and CO cloud distances can be made by
integrating the {\hi} line profiles towards the stars from zero velocity to the
velocity where an {\hi} column corresponding to ${\rm E}_{{\rm B}-{\rm V}}
\approx 1$ is found. For the WR~5 and HD~17603 sight lines, reddening implies
an {\hi} column of $N \sim4.5 \times 10^{21}$~{\cmii}. Our {\hi} data shows
that this column density is reached when the line profiles are integrated from
zero velocity out to $\approx -50$~{\kms}, consistent with the radial velocity
of {\cog}. 

\subsection{The B Stars}

In addition to the Wolf-Rayet and OIb(f) star discussed above, there are eight
B stars within $\sim 0\fdg5$ of the tip of {\hig}. Among them is EO~Per, a
rapidly varying B0 supergiant. A B0 supergiant has a mass of roughly 30~{\msun}
and so was approximately type O7.5 while on the main sequence. According to
\citet{how89}, such a star has a wind terminal velocity of 2300~{\kms},
slightly lower than the lower bound quoted for both WR~5 and HD~17603, and a
mass loss rate of $10^{-6.7}$~{\msunyr}, an order of magnitude smaller than the
lower bound for the other two stars. At present this star is likely to have a
higher mass loss rate but much lower wind terminal velocity than WR~5 and
HD~17603, as well as a lower ionizing flux.

The remaining seven candidates have very limited observational data upon which
to base determination of their type and distance. If we assume they are located
at $D \approx 2$~kpc and lie behind extinction of ${\rm A}_{\rm V} \approx 3$,
then all these stars have photometry roughly consistent with early B type stars
on the main sequence, although we cannot rule out the possibility that some
are actually O type stars.

\section{The Interstellar Environment of WR~5 and HD~17603}\label{sec:environ}

Based on low resolution ($9\arcmin$) observations, \citep{arn92} proposed that
WR~5 is associated with an ovoidal {\hi} deficiency, interpreted as a cavity,
at velocities around $-14$~{\kms}. We note that integrating out to $-14$~{\kms}
in the {\hi} line yields a column density only $\sim1.5 \times
10^{21}$~{\cmii}, corresponding to ${\rm E}_{{\rm B}-{\rm V}} \approx 0.3$,
which is too small to be consistent with the observed reddening to the two
stars. WR~5 and HD~17603 are well beyond the distance corresponding to that gas
velocity and thus cannot be associated with the ovoidal structure. Our high
resolution ($1\arcmin$) observations show that the {\hi} distribution is highly
structured in both morphology and velocity, and the ``cavity'' is by no means
clearly distinguishable. In addition, regions of depressed {\hi} brightness
within this ``cavity'' correspond in position and velocity with much of the CO
emission seen around $-14$~{\kms}, which suggests that the relative lack of
{\hi} emission is due at least in part to the hydrogen being in molecular
rather than atomic form. 
 
For the velocity range within which we associate WR~5 and HD~17603 ($-53.5 \la
v \la -42.0~{\kms}$), our {\hi} data do not show any convincing evidence for an
{\hi} shell surrounding a cavity. However, it is possible that a gas shell does
exist but is not clearly visible in the {\hi} observations because of its low
contrast or because parts of it are molecular. The clear
anti-correlation between the CO clouds and the {\hi} brightness distribution in
the $-53.5$ to $-42.0~{\kms}$ velocity range and the relative lack of CO
emission within $\sim40\arcmin$ of the stars lend some support to the latter
possibility.

\citet{mar96} cited evidence from IRAS Skyflux images for a $1\fdg35$-diameter
dust shell around WR~5. Such a shell is not obvious in the HIRES processed 12,
25, 60, and $100~{\mu}$m mosaics. To investigate further, we used the 60 and
$100~{\mu}$m data to produce dust temperature and optical depth images. The
scattered dust concentrations interpreted by Marston as a shell correlate much
better with the velocity-integrated CO data than with the velocity-integrated
{\hi} data, some with CO emission in the velocity range around $-14$~{\kms} and
others with the CO around $-48$~{\kms}. This suggests that what was interpreted
as a shell is in fact the result of projection effects. 

We inspected {\ha} data for this region from the Virginia Tech Spectral-line
Survey \citep{den98} and the Wisconsin H-Alpha Mapper 
\citep{haf03} Northern Sky Survey. The VTSS has an angular resolution of
$\sim 1\farcm6$ but does not provide velocity information. The WHAM {\ha} data
has a velocity resolution of $12$~{\kms} but a lower angular resolution
$\sim 1\degr$. The VTSS data suggest that WR~5 and HD~17603 both lie within a
region relatively devoid of {\ha} emission. The morphology of this ``void'' and
thus its possible association with the stars is uncertain because a comparison
of the VTSS image with velocity-integrated CO emission suggests that absorption
by the dust in molecular clouds strongly influences the observed {\ha}
brightness distribution. 

We conclude that there is at present no conclusive evidence for an {\hi} or
dust shell/cavity structure around WR~5 or HD~17603, although the stars appear
to lie in regions relatively free of molecular, neutral atomic, and ionized
gas. Any gas shell surrounding this region would have to be of low contrast or
be rather inhomogeneous in its composition and structure: the latter implies a
``leaky'' shell within which a hot wind bubble might not be maintained. 

\section{Discussion}\label{sec:discuss}

The morphological relationships between the {\hi} tails, the CO clouds, and the
stars raises the question of whether all these are physically related. The
possibility of a simple coincidence cannot be ruled out, but an explanation for
the morphology and velocity structure of the {\hi} tails would still be needed.
In the remainder of this paper we investigate the possibility that winds from
massive stars could be responsible for driving long narrow {\hi} tails from
compact molecular clouds in their vicinity.

A viable physical model of {\hig} must be consistent with all of the
observational data we have presented above. Obviously, we need a stellar
mechanism for converting molecular to neutral atomic hydrogen and accelerating
the {\hi} away from the molecular clumps in long narrow tails at a
velocity of $\sim 5 - 20$~{\kms}. WR~5 and/or HD~17603 must be able to supply
momentum and energy at the rates required over a time comparable to the
dynamical timescale. The molecular clumps must also be able to survive within
the environment of the star(s) for this time.

One question that immediately arises is the association of {\hi} tails with
only some of the CO clouds in the $-44 \la v \la -50$~{\kms} velocity range.
Standard Galactic rotation curves predict a slope of the distance$-$velocity
relationship at this location in the Galaxy of $\sim 150$~{\pckms}. The
above velocity range thus potentially covers a line of sight range of
$\sim 1$~{\kpc} and the velocity resolution of our CO data ($0.98$~{\kms})
corresponds to $\sim 150$~{\pc}. The empirical rotation curve of \citet{bra93}
suggests the slope may be even steeper. Given the potentially very wide range
of distances for the CO clouds, only a small subset of the observed clouds may
be within the range of influence of the tail-driving source(s). Additionally,
{\hi} flows from other CO clouds may be subject to confusion if they have
relatively low radial velocities.

\subsection{Kinematics and Timescales}\label{sec:kin_time}

Determining the mass of extended {\hi} structures is always subject to wide
uncertainties, not least because of confusion with unrelated emission. This is
particularly acute for structures having large velocity dispersions. We
determined the mass of {\hig} by first removing an estimate of the background
{\hi} emission on a channel-by-channel basis. This background was calculated by
a linear interpolation anchored to velocity channels bracketing the velocity
range of {\hig} which were judged to be free of emission from the latter. The
background subtracted cube was then integrated over a polygonal region
enclosing {\hig}. The solid angle of this region was 1.1 square degrees, and
the mean {\hi} column density derived was $N = 1.2\times 10^{20}\cmii$. The
total {\hi} mass is $330 D^{2}$~{\msun} and the total mass, assuming a hydrogen
mass fraction of 0.77, is $M = 425 {\rm D}^{2}$~{\msun}. The uncertainty in the
mass estimate is at least $25\%$. The physical length of {\hig} projected onto
the plane of the sky is $\sim 30 D$~pc.

Assuming $D = 2$~kpc for {\hig}, this yields a total mass of the structure of
$M = 1700\, {\msun} = 3.38\times 10^{36}$~g. Assuming a systemic radial
velocity of the mass reservoir of $-47$~{\kms}, the linear momentum contained
in {\hig} is $P = P_{0}/\cos(i) = 1.05\times 10^{42} {D}^2/\cos(i) =
4.20\times 10^{42}/\cos(i)$~{\gcms}, where we have assumed that the space
velocity of the flow can be deprojected from the radial velocity by using an
inclination angle to the line of sight $i$. The mass-weighted mean outflow
velocity is then $12.4/\cos(i)$~{\kms}, and the kinetic energy is $E =
E_{0}/\cos^{2}(i) = 6.67\times 10^{47} (D/\cos(i))^{2} = 2.67\times
10^{48}/\cos^{2}(i)$~erg. The driving source(s) presumably have injected
momentum and energy isotropically into the ISM. The observed opening angle of
{\hig} is $\theta \approx 20\degr$: assuming that the three dimensional
structure is conical and viewed at an inclination $i$, the true opening angle
is $\theta \sin(i)$ , and {\hig} subtends a fraction $f = (1-\cos(\theta
\sin(i)/2))/2$ of a sphere. Therefore, $f \leq 7.6\times 10^{-3}$ ($\leq
0.1$~ster), the maximum value occurring when $i = 90\degr$ (when the outflow
axis lies in the plane of the sky). Over the lifetime of {\hig} (discussed
below), the driving source(s) have supplied a total (isotropic) momentum and
kinetic energy of, respectively, $P_{*} = P/(f \epsilon_{P})$ and $E_{*} = E/(f
\epsilon_{E})$, where the $\epsilon$ factors represent the efficiency of the
transfer of momentum and energy to the {\hi}. The minimum value of $P_{*}$ and
$E_{*}$ is obtained by minimizing the following two expressions with respect to
$i$: $$\frac{\epsilon_{P} P_{*}}{P_{0}} = \frac{1}{f \cos(i)} =
\frac{2}{\left(1 - \cos\left(\frac{\theta \sin(i)}{2}\right)\right) \cos(i)}$$
$$\frac{\epsilon_{E} E_{*}}{E_{0}} = \frac{1}{f \cos^{2}(i)} = \frac{2}{\left(1
 - \cos\left(\frac{\theta \sin(i)}{2}\right)\right) \cos^{2}(i)}$$

For $\theta = 20\degr$, the minimum values are $P_{*, min} = 342
P_{0}/\epsilon_{P} = 1.43\times 10^{45}/\epsilon_{P}$~{\gcms} (when $i \approx
55\degr$), and $E_{*, min} = 526 E_{0}/\epsilon_{E} = 1.40\times
10^{51}/\epsilon_{E}$~erg (when $i \approx 45\degr$). A simple conical outflow
model yields for the flow timescale $\tau = \tau_{0}/\tan(i) = 2.36\times
10^{6}\, D/\tan(i) \approx 4.72\times 10^{6}/\tan(i)$~yr, where we have used
the mass-weighted mean outflow radial velocity $12.4$~{\kms} and the
mass-weighted mean angular extent $1\fdg72$. We have noted above the presence
of a strong uniform radial velocity gradient along the {\hi} tails. If this
were a manifestation of velocity-sorting rather than continuous acceleration,
$\tau$ might be more accurately estimated by using the maximum outflow radial
velocity observed ($\sim 20$~{\kms}) rather than the mean velocity, resulting
in a factor of 2 decrease in $\tau$. We will argue below that the velocity
gradient is more likely to be the result of a steady acceleration of the tails,
in which case the use of the mean velocity is approximately correct.

Using the dynamical timescale we can estimate the rate at which the driving
source(s) deliver (isotropically) momentum and kinetic energy, respectively
$\dot{P_{*}}$ and $\dot{E_{*}}$. The minimum values of $\dot{P_{*}}$ and
$\dot{E_{*}}$ are obtained by minimizing the two following expressions with
respect to $i$:
$$\frac{\epsilon_{P} \dot{P_{*}} \tau_{0}}{P_{0}} = \frac{\sin(i)}{f
\cos^{2}(i)} = \frac{2 \sin(i)} {\left(1 - \cos\left(\frac{\theta
\sin(i)}{2}\right)\right) \cos^{2}(i)}$$

$$\frac{\epsilon_{E} \dot{E_{*}} \tau_{0}}{E_{0}} = \frac{\sin(i)}{f
\cos^{3}(i)} = \frac{2 \sin(i)}{\left(1 - cos\left(\frac{\theta
sin(i)}{2}\right)\right) \cos^{3}(i)}$$

We find $\dot{P}_{*, min} = 341 P_{0}/(\epsilon_{P} \tau_{0}) = 9.6\times
10^{30}/\epsilon_{P}$~{\gcmss} (when $i = 35\degr$), and $\dot{E}_{*, min} =
405 E_{0}/(\epsilon_{E} \tau_{0}) = 7.2\times 10^{36}/\epsilon_{E}$~{\ergs}
(when $i = 30\degr$). Figure~\ref{fig:dep_on_i} shows plots of the dependence
of $P_{*}$, $\dot{P_{*}}$, $E_{*}$, and $\dot{E_{*}}$ on $i$ for the simple
conical flow model. The uncertainty in the values of the kinematic and
energetic parameters is dominated by our lack of knowledge of the inclination
angle, affecting in particular the value of the kinematic timescale. In our
discussions below we will assume that $i$ most likely lies within the range
$30\degr \la i \la 60\degr$. Our estimates of the physical parameters of the
{\hi} tails are summarized in Table~\ref{tbl:tail_param}.
 
\subsection{Mass-loaded Flow Model}\label{sec:mass-loaded}

The lowest velocity shift discernable between the CO and {\hi}, limited by
background confusion, is $\sim 5$~{\kms}, which is roughly comparable to the 
sound speed in neutral atomic hydrogen. The systematic radial velocity gradient
of {\hig} could be interpreted in terms of curvature along the line of sight or
of expansion. However, if {\hig} is the result of a flow from the CO clouds, it
is reasonable to assume that the {\hi} filaments are radially directed away
from the driving source and are thus approximately linear. The velocity
gradient would then reflect a continuous acceleration of gas along the
filaments, leading to the picture of a flow embedded within an extended wind
region. This picture of long flowing tails from dense clouds driven by an
external wind is similar to the situation described analytically by
\citet{dys93} and numerically by \citet{fal02} of the quasi-steady hydrodynamic
mixing and entrainment of dense clump gas by a diffuse subsonic or transsonic
wind. We propose that the neutral atomic tails are created in the dissociation
and entrainment of molecular clouds enveloped within a strong stellar wind or
winds. The wind ablates and entrains cloud gas and accelerates it away from the
cloud in long narrow tails as a mass-loaded flow. Given the positional
associations noted above, the source of the wind(s) is likely to be in the
vicinity of WR~5 and HD~17603. 

We note that it is very unlikely that these {\hi} filaments are cometary tails
caused by shadowing of radiation in a low-density {\hii} region. The flaring of
the tails away from the axis of symmetry with increasing distance from the apex
is inconsistent with {\hi} observations \citep{mor96, hey96} and models
\citep{ber90, can98, pav01} of cometary clouds. The dynamical timescale of
{\hig}, a few million years, is far longer than the timescale for complete
photoionization of the tails \citep{pav01}. This leaves the wind-driven model,
but in order to establish the plausibility of this model there are several
questions which must be addressed. Can the molecular clouds survive a few 
million years inside a stellar wind region? What must be the conditions inside
this region in order to account for the observed cloud mass loss rate and
entrained gas acceleration, and how realistic are they? Are the energetics
plausible? What are the mechanisms
by which the entrained gas may be dissociated? We address these questions in
the following sections.

\subsubsection{Cloud Survival}

For the proposed model to be feasible, the molecular clouds must survive their
initial impact with the stellar wind, before settling down to a quasi-steady
entrainment phase within the wind. \citet{kle94} have performed an analytic and
numerical study of the interaction between a fast, steady, planar shock and a
dense isothermal, nonmagnetic, spherical, cold cloud. More recently,
\citet{pol02} have extended numerical studies to the case of shocks
encountering multiple clouds in which cloud-cloud interactions can oocur. 
As summarized in both these papers (we follow the formulation of \citet{pol02}
in this paper), a key timescale is the
``cloud-crushing time'' $t_{cc}$, which is the time
required for the internally transmitted shock to cross the cloud, $t_{cc}
\equiv {2 a_0}/v_{cs}$. Here $a_0$ is the initial cloud radius and
$v_{cs}$, the forward velocity of the shock through the cloud, is
$$v_{cs}\approx {{v_s}\over{\chi^{1/2}}}{(F_{c1}F_{st})}^{1/2}$$ 
where $F_{c1}$ is the ratio of the pressure just behind
the cloud shock to that at the stagnation point
and $F_{st}$ is the
ratio of the pressure at the stagnation point to that in the far upstream
wind. Both these quantities have values of roughly unity \citep{kle94, pol02}.
The velocity of the external shock is $v_{s}$ and the density contrast
$\chi = \rho_{c0} / \rho_{i0}$ is
the ratio of the initial cloud density to that of the external 
medium. In the numerical simulations of \citet{kle94} and \citet{pol02},
the cloud destruction time $t_{dest}$ (defined as the time at which the mass of
the surviving cloud core is reduced to $\sim 0.5$ of its initial value) was
typically found to be $t_{dest} \sim 2 t_{cc}$. The numerical simulations
of \citet{mac94} suggest values of $t_{dest} \sim 4 t_{cc}$ with the
inclusion of dynamically significant magnetic fields.

The molecular clouds at the apex of the {\hi} tails have a total mass $\sim
440$~{\msun}. Their current radii are $a \approx 2$~{\pc} and they have H$_2$
number densities $n_{{\rm H}_2} \approx 150$~{\cmiii}. At their present stage
of evolution, the clouds have gravitational virial parameters $\alpha \equiv 5
\sigma^2 a / {\rm G} m_c \approx$ 80--150, indicating that they are not bound
by their self-gravity ($\sigma$ is the observed cloud velocity dispersion,
$\sim 3.3$~{\kms}, and $m_c$ is the cloud mass).
If we assume that the total mass of the {\hi} tails was originally in the
molecular clouds, we find that the clouds have been reduced from their initial
mass by a factor of about 5: according to the formal definition of $t_{dest}$
above, these clouds would be classified as largely destroyed.

If we set $t_{dest} \equiv \zeta t_{cc}$, the cloud destruction time is
$$\frac{t_{\rm dest}}{\yr} \sim 1\times 10^6 \zeta \chi^{1/2}
\left(\frac{a_0}{\pc}\right) \left(\frac{v_s}{\kms}\right)^{-1}.$$ 
For shock velocities $v_s \sim
100$~{\kms} and an initial density contrast $\chi > 100$ 
(with $a_o$ a few parsec and $2 \la \zeta \la 4$), the clouds will survive for
times of order
a few $10^6$~{\yr}, long enough for them to become enveloped by the wind.
We conclude that the molecular clouds can survive the passage of the initial
shock, after which they are gradually destroyed while mass-loading the
surrounding flow.

\subsubsection{Conditions in the Mass-loading Region}

The mass-loading of flows by embedded clouds has been discussed in a series of
papers by Dyson and collaborators
\citep{har86,dys87,cha88,art93,dys93,art94,dys94}. For an initially
pressure-confined cloud which becomes immersed in a wind flow, the pressure
gradients across the cloud surface cause a lateral expansion of the cloud which
mixes cloud gas into the wind. \citet{har86} show that the mass loss rate from
the cloud is
\begin{center}
$\begin{array}{l l l l}
\dot{m}_c & = & \alpha M^{4/3} (m_c C_c)^{2/3} (\rho_w v_w)^{1/3} & {\rm
for}\ M \leq 1 \\
\dot{m}_c & = & \alpha (m_c C_c)^{2/3} (\rho_w v_w)^{1/3} & {\rm for}\ M
\geq 1
\end{array}$ 
\end{center}
\noindent
where $C_c$ is the isothermal sound speed in the cloud, $\rho_w$ and $v_w$ are
the wind density and velocity, and $M = v_w / C_w$ is the Mach number in the
wind. In this expression, $\alpha$ is a constant of order unity for which
\citet{har86} provisionally assign a value $1/2$. Assuming that the wind and
cloud parameters (except $m_c$) are roughly constant over the cloud lifetime, 
the cloud mass loss rate averaged over the entire cloud lifetime is
$\langle\dot{m}_c\rangle \approx \dot{m}_{c0}/3$, where $\dot{m}_{c0}$ is the
maximum mass loss rate corresponding to the initial mass $m_{c0}$ of the cloud.
Therefore,
$$\frac{\langle\dot{m}_c \rangle}{\gs} \approx 3.41 \times 10^{18} M^{4/3}
\left(\frac{m_{c0}}{{\msun}}\right)^{2/3} \left(\frac{C_c}{\kms}\right)^{2/3}
\left(\frac{n_w}{{\cmiii}}\right)^{1/3} \left(\frac{v_w}{{\kms}}\right)^{1/3}$$
where $n_w$ is the number density of the cloud (we have assumed a mean mass per
particle of $\mu = 2.17 \times 10^{-24}$~{\g}).

For simplicity, we assume that the initial cloud mass (the current molecular
cloud mass plus the mass in the {\hi} tails) was contained within a single
cloud, thus $m_{c0} \sim 2100$~{\msun}. The relative fraction of the initial
mass in the tails and the clouds indicates that $\tau \approx 0.4 t_{lt}$,
where $t_{lt}$ is the cloud lifetime predicted by integrating the cloud mass
loss rate. We thus appear to be observing the clouds near
the half-way point in time of their destruction process. 
The mass loss rate averaged over this time period (the dynamical
timescale $\tau$) is $\langle \dot{m}_c(\tau) \rangle \sim 2 \dot{m}_{c0} /3$,
roughly twice $\langle \dot{m}_c \rangle$. Using our value of $m_{c0}$, we
obtain
$$\frac{\langle \dot{m}_c(\tau) \rangle}{\gs} \approx 1.13 \times 10^{21}
\left(\frac{C_c}{\kms}\right)^{2/3} \left(\frac{n_w}{{\cmiii}}\right)^{1/3}
\left(\frac{v_w}{\kms}\right)^{1/3}$$
where we have assumed $v_w \approx C_w$ (i.e. $M \approx 1$). We have observed
a value $\langle\dot{m}_c(\tau)\rangle \approx\ $1.3  -- 3.9$ \times
10^{22}$~{\gs} (for $30\degr \leq i \leq 60\degr$), which implies that we can
match the theoretically predicted mass loss rate if
$$1.5 \times 10^3 \la
\left(\frac{C_c}{\kms}\right)^{2} \frac{n_w}{\cmiii} \frac{v_w}{\kms} \la 4.2
\times 10^4.$$

In the theory of \citet{har86}, the isothermal sound speed $C_c$ is better
identified as the characteristic speed at which the dense cloud can supply mass
to the wind-cloud mixing region. Thus, we will assume that the effective value
of $C_c$ is $\sigma \approx 3.3$~{\kms}. Then, the above condition becomes
$$140
\la \frac{n_w}{\cmiii} \frac{v_w}{\kms} \la 3800$$ and has a value of 
$\approx 730$ when $i = 45\degr$. The wind velocity experienced
by the dense clouds cannot be less than the maximum velocity observed in the
{\hi} tails ($\sim 25$--40~{\kms}, depending on $i$). The maximum possible wind
speed is however much larger: the winds of massive main sequence stars can have
terminal velocities $\approx 3000$~{\kms} (cf. Table~\ref{tbl:WR5_HD17603}). If
$v_w \approx 3000$~{\kms}, the mass-loading model predicts $0.05 \la n_w/{\cmiii} \la 1.4$. Such large values of $n_w$ rule
out the possibility that the tails are produced in the mass-loading of clouds
immersed in a freely-expanding bare stellar wind. Direct momentum-driving of
the tails by a wind is also ruled out: the total isotropic linear momentum
supplied by the wind of a 40--85~{\msun} star during its main sequence lifetime
of a few $10^{6}$~{\yr} is approximately $10^{42}$ -- $10^{43}$~{\gcms}
\citep{sch96}, which is at least two orders of magnitude smaller than
$P_{*, min}$.  

The isotropic momentum flux predicted by the mass-loading model is
$$\dot{P} = 4 \pi r^2 \rho_w {v_w}^2 = (9.2-83. \times 10^{24}) v_w~{\gcmss},$$
where substituting our adopted range of values of $n_w v_w$ leaves only one
factor of $v_w$ (the separation between the driving source(s) and the CO clouds
is $r = r_0 /\sin\theta$, where we have assumed $r_0 = 25$~{\pc}).
Equating $\dot{P}$ with our observed value of $9.6 \times 10^{30}
\la \dot{P}_* \la 1.7 \times 10^{31}$~{\gcmss}, we see that wind speeds in the
mass-loading region greater than $v_w \sim 10^6$~{\cms} ($\sim 10$~{\kms})
would be sufficient to account for the momentum flux. 
Given that velocities in {\hig}
may be as high as $\sim 40$~{\kms}, $v_w$ must be significantly higher than the
limit of $\sim 10$~{\kms} we have obtained. We note that the predicted value of
$n_w v_w$ in the mass-loading theory is very sensitive to the value of the
``mixing efficiency'' $\alpha$ ($n_w v_w \propto \alpha^{-3}$). The mixing
efficiency may well be smaller \citep{can91,art97}. Lowering the value of the
mixing efficiency by a factor of 2 or 3 would increase the required wind speeds
in the mass-loading region by a factor of $\sim 10$, to $\sim 100$~{\kms}.
Using this higher value, the mass-loading theory predicts a wind density in the
mass-loading region of $n_w \sim 1$--40~{\cmiii}.

We conclude that the mass-loading scenario is workable in the context of the
main sequence stellar winds from massive O stars if the
fast low-density stellar wind is significantly loaded with mass and greatly
slowed before interacting with the dense cloud. The low inferred wind velocity
is consistent with the relative narrowness of the {\hi} tails as predicted
theoretically \citep{dys93} and found numerically \citep{fal02} for the case of
mass loading by a mildly supersonic (transsonic) wind. A related point concerns
the apparent complete mixing of mass injected into the wind before encountering
the CO clouds: in contrast the {\hi} filaments clearly do not completely mix
over a large distance. Mixing may be greatly enhanced in a highly
inhomogeneous medium having numerous clumps which are close enough togther for
wind-driven interactions to occur \citep{pol02, gar02}. In contrast, the CO
clouds associated with the {\hi} filaments are likely to be the dominant mass
concentrations in their vicinity and so are not subject to mixing driven by
cloud$-$cloud interactions.

Our model still
requires that the wind transfer very large amounts of kinetic energy to the 
{\hi} tails: several times $10^{36}$~{\ergs}, which is
the same order of magnitude as the mechanical luminosity of the main sequence
wind of a massive O star. This is a severe constraint,
since in the basic
theory of adiabatic stellar wind bubbles, only a tiny fraction, $\sim 1\%$, of
the energy is in kinetic form within the bubble, the bulk being in the thermal
energy of hot shocked gas and in the kinetic energy of the swept-up shell.
Therefore a very small fraction of the energy is available to drive
mass-loading of clouds trapped inside.
 
Much higher efficiencies of energy conversion into kinetic form are possible if
the stellar wind is itself significantly mass-loaded. \citet{art94} have shown
that mass-loading of an initially highly supersonic wind by mass sources
distributed smoothly throughout the region around the star can reduce the Mach
number of the wind without having the wind going through a global shock near
the star. If the mass loading is sufficiently heavy
($m_{\mathrm{load}}/m_{\mathrm{wind}} > 10$, where $m_{\mathrm{load}}$ is the
mass loaded into the wind and $m_{\mathrm{wind}}$ is the mass of the wind
injected by the star), a large fraction of the stellar wind energy ($> 50\%$)
will remain in the form of kinetic energy within the wind region \citep{pit01}.
Therefore, stellar winds which are heavily ``pre-loaded'' with mass may be able
to provide the kinetic energy required to drive flows from larger-scale mass
concentrations. It also may potentially explain the low wind velocities and
high wind densities that we infer. In this regard, a scaled-up version of the
scenario modelled numerically by \citet{gar02}, the mass-loading of the
wind from the Trapezium cluster by propylds, may be relevant. \citet{gar02}
showed that mass-loading could have very significant dynamical
effects on the wind, including increased wind density and low Mach numbers at
large radii. 
However, even with a high efficiency of 
momentum transfer to the {\hi} filaments, the momentum input of the winds of
WR~5 and HD~17603 integrated is too low by at least an order of magnitude.
It would seem then that our scenario requires the wind input from a group of 
many more O stars in addition to these two.

We have not been able to identify such a grouping of stars. Since massive O
stars form in OB clusters, short-lived stars 
such as WR~5 and HD~17603 do not have time to travel far from their
birthplaces (unless they are runaway stars ejected from a binary during the 
supernova explosion of the other component). Although it is not satisfactory to
have to posit the existence of an unseen group of OB stars, the apparent
close proximity of WR~5 and HD~17603 to each other makes this idea more
plausible. Given the effects of visual extinction (and possible anomalous
extinction) in a direction towards a major spiral arm, and the difficulties
of finding, classifying, and obtaining distance estimates, it is perhaps not
unlikely that such a grouping of stars could have escaped detection. Given
the dynamical timescale of the {\hi} filaments, it is quite possible that the
parent molecular cloud in which the group formed has been largely destroyed by
now, and some of its members no longer exist after having gone supernova (we
could speculate that supernova explosions might have contributed to the flow
driving). Such a remnant grouping of stars might be very difficult to identify.

\subsubsection{Dissociation}

There are two plausible mechanisms for dissociating the molecular gas which
forms the {\hi} tails: photodissociation and shock dissociation. Stars like
WR~5 and HD~17603 produce very large fluxes
of ionizing photons ($\gtrsim 10^{49}$~{\pers}), and thus the ionized skins
of the molecular clouds may be observable in {\ha}. The {\ha} brightness can
be estimated using, e.g., equation $13$ of \citet{lop01}, where we assume a
cloud radius 2~{\pc}, a ratio of ionized skin thickness to cloud radius of
$0.1$, and a source of ionizing photons $\sim 7.5\times 10^{49}$~{\pers} located
$25$~{\pc} from the cloud. The latter was estimated using equation $1$ in
\citet{gre99}. The resulting {\ha} surface brightness observed from the Earth,
assuming no extinction, corresponds to a photon specific flux of
$\sim 3.6\times 10^{7}$~{\psf}. This is comparable to the surface brightnesses
of the fainter
photoevaporating cometary knots observed in the Helix nebula by \citet{ode00}.
If the
clouds suffer a similar extinction as WR~5, the {\ha} emission will be
further attenuated by a factor of $\approx 10$. The large (${\rm A}_{\rm V}
\sim 3$) and patchy visual obscuration may make detection of {\ha} emission
from the clouds difficult (see discussion in section~\ref{sec:environ}). 
There is no evidence
for radio continuum emission from ionized hydrogen from the CGPS 1420~{\mhz}
continuum data. Any {\hii} region must thus be of very low radio surface
brightness: a 100~{\pc} path length through a 1~{\cmiii} {\hii} region would
produce only a few tenths of a kelvin brightness temperature, which could not easily
be distinguished from the general Galactic background of several kelvin.

Hot massive stars produce photons capable of dissociating molecular hydrogen at
rates comparable to the ionizing photon rates \citep{dia98}. A rough evaluation
of the possibility that the tails are photodissociated can be made by comparing
the photodissociating rate into the solid angle of the tails, $p S_D f$, to the
dissociation rate implied by our observations. $S_D \sim 10^{49}$~{\pers} is
a typical stellar photodissociating photon rate for a very massive O star,
$p \approx 0.15$ is the probability
that an absorption will lead to dissociation \citep{dia98}, and $f$ is the
fractional solid angle subtended by the tails. We find that the two rates are
roughly equal at $\sim 10^{60}$ dissociations when integrated over a period of
a few $10^6$~{\yr}. Photodissociation thus appears to be a plausible
dissociation mechanism for the tails.

The other possibility is that shocks driven into the molecular clouds by the
stellar wind(s) may be capable of dissociating the H$_2$. J-type 
shocks, in which the shock velocity $v_{cs}$ is sufficiently high to preclude
the
existence of a magnetic precursor, are the most effective at dissociating
molecular gas. In a strong J-type shock with $v_{cs} \ga 50$~{\kms} moving into
molecular gas with a pre-shock H$_2$ number density $n \sim 100$~{\cmiii},
about $90\%$ of the H$_2$ will be dissociated \citep{hol80}. 
For $\chi \sim
10^{2}$ and a wind velocity $v_w$ of the order of a few hundred {\kms}, shock
velocities will be of the order of $40$--$50$~{\kms}, which is sufficient to
result in very significant dissociation. The energy required to shock
dissociate the molecular gas is roughly an order of magnitude smaller than that
to accelerate it to the velocity of the tail. However, shock velocities of
this magnitude may reduce the cloud destruction time to values
$< 10^{6}$~{\yr}, which is a problem for the scenario we are proposing.

We conclude that photodissociation is the most workable mechanism for 
dissociating the molecular gas swept into the {\hi} filaments. 

\section{Summary and Conclusions}
\label{sec:conclude}

We have discovered an unusual system of {\hi} ``tails'' comprised of a number
of linear or quasi-linear thin filaments which, if projected along their
length, appear to have a common origin. The feature, which we dub
{\higfull}, has dimensions of $1\fdg8 \times 0\fdg8 \times 15$~{\kms}
(length $\times$ width $\times$ velocity), although its width varies along
its length.

Located at one end of the filaments is a chain of molecular clouds that
lie at a distance of approximately 2~{\kpc}. Despite the molecular material
being displaced from the atomic by $\sim 10 - 30$~{\kms}, we propose that the
molecular clouds and {\hi} feature are located at the same distance and
are dynamically related. The lack of a traditional head-tail morphology,
and the lack of any evidence of ionization along the leading edge of the
molecular clouds, lead us to reject the possibility that this system is
simply a cometary cloud; instead, we propose that the {\hi} feature is
made up of material dissociated from the molecular clouds 
and entrained in
the outflowing wind as a mass-loaded flow. 

Given the velocity difference between the atomic and molecular material,
we estimate the dynamical age of the system to be on the order of a few 
$10^6$ years. The velocity gradient observed along the {\hi} feature is,
in our model, explained by continuous acceleration of the
entrained material by the stellar wind. 

A search for driving sources for the {\hi} filaments identified the
Wolf-Rayet star WR~5 and the Of star HD~17603 as candidates. Both these stars
are located
near the apparent origin of the filaments, are approximately at the same
distance as the molecular clouds, and have main sequence lifetimes
comparable to the dynamical age of the system. However, even with the high 
efficiency of momentum transfer predicted for a heavily mass-loaded wind,
WR~5 and HD~17603 are incapable on their own to drive the filaments: we are
forced to speculate that other, unidentified stars and/or old supernova
explosions must also have been involved. We cannot of course discount the
possibility that the positional associations between {\hi}, CO, and stars
are simply accidental.
But, assuming that stellar winds (or explosions) have shaped the gaseous
structures we have observed, we suggest that anisotropic {\hi}
structures similar to {\hig} should be searched for in the environments around
massive stars (and supernova remnants). 

\acknowledgements{We thank Tom Landecker for his support and encouragement, and
the referee for comments that led to a number of key improvements in this paper.
The Dominion Radio Astrophysical Observatory Synthesis
Telescope is operated as a national facility by the National Research Council
of Canada. The Canadian Galactic Plane Survey is a Canadian project with
international partners, and is supported by grants from the Natural Sciences
and Engineering Research Council of Canada. Guest User, Canadian Astronomy
Data Centre, which is operated by the Herzberg Institute of Astrophysics,
National Research Council of Canada. Data from the Canadian Galactic
Plane Survey is publicly available through the facilities of the Canadian
Astronomy Data Centre (http://cadc.hia.nrc.ca).}

\clearpage

\begin{deluxetable}{l l l}
\tablewidth{0pt}
\tablecaption{Parameters of Relevant CGPS Data
\label{tbl:CGPS_param}}
\tablehead{
\colhead{Data Product} & \colhead{Parameter} & \colhead{Value}
}
\startdata
{\hi} 21~cm line & Spectral resolution & 1.32~{\kms} \\
 & Channel width & 0.82~{\kms} \\
 & Angular resolution & 64{\arcsec} \\
\hline
 & &  \\
{\coi} line & Spectral resolution & 0.98~{\kms} \\
 & Channel width & 0.82~{\kms} \\
 & Angular resolution & 100{\arcsec} \\
\enddata
\end{deluxetable}

\clearpage

\begin{deluxetable}{l l l l l}
\tablewidth{0pt}
\tablecaption{All O, B and Wolf-Rayet stars
  within $0.5^\circ$ of the tip of \hig. $^{[a]}$
\label{tbl:stars}}
\tablehead{\colhead{Name} & \colhead{Spectral type} &
  \colhead{G.long.} & \colhead{G.lat.} & \colhead{Comments}}
\startdata
WR 5 & WC6 $^{[b]}$ & 138.87 & $-2.15$ & \\
HD 17603 & O7.5Ib(f) $^{[c]}$ & 138.77 & $-2.08$ & \\
EO Per & B0Iab:e & 138.50 & $-1.89$ &  $\rceil$ Listed RA/dec differ slightly
\\
CDS 314 & B & 138.50 & $-1.89$ & $\rfloor$ for these two stars  \\
BD+56\degr 722 & B & 138.57 & $-2.14$ & \\
ALS 7516 & B & 138.80 & $-2.08$ & \\
ALS 7523 & B & 138.87 & $-2.04$ & \\
ZZ Per & B & 138.84 & $-2.19$ & \\
ALS 7453 & B & 138.22 & $-2.32$ & \\
ALS 7470 & B & 138.16 & $-1.75$ & \\
\enddata
\tablenotetext{[a]}{Compiled using the SIMBAD database. Unless
  otherwise noted the information in this table is that listed in SIMBAD.}
\tablenotetext{[b]}{\cite{van01}.}
\tablenotetext{[c]}{\cite{HD17603type}.}
\end{deluxetable}

\clearpage

\begin{deluxetable}{l l l}
\tablewidth{0pt}
\tablecaption{Wolf-Rayet 5 and HD 17603 \label{tbl:WR5_HD17603}}
\tablehead{\colhead{Parameter} & \colhead{WR5} & \colhead{HD 17603}}
\startdata
Spectral type & WC6 $^{[a]}$ & O7.5Ib(f) $^{[d]}$ \\
v$_\infty$ & 2365 \kms\ $^{[b]}$ &  2500 \kms\ $^{[e]}$ \\
\mdot & 10$^{-4.3}$ \msunyr\ $^{[c]}$ & 10$^{-6.79}$ \msunyr\ $^{[f]}$ \\
$L_w$ & $8.9 \times 10^{37} {\rm erg\ s}^{-1}$ & - \\
\\
MS spectral type $^{[0]}$ & O5 & O5 \\
MS v$_\infty$ & 2800 \kms & 2800 \kms \\
MS \mdot & 10$^{-5.7}$ \msunyr & 10$^{-5.7}$ \msunyr \\
MS $L_w$ & $4.5 \times 10^{36}$ erg s$^{-1}$ & $4.5 \times 10^{36}$ erg s$^{-1}$ \\
MS lifetime & 3.7 Myr & 3.7 Myr \\
MS wind energy & $5 \times 10^{50}$ erg & $5 \times 10^{50}$ erg \\
\enddata
\tablenotetext{[0]}{The main sequence spectral types were estimated from
       the evolutionary tracks by \citet{maeder90}.
        WR5 would have had a ZAMS mass of 40 \msunyr\ or more
        which corresponds to a main sequence spectral type of O6.5 or
        earlier. The type O5 was chosen to give a rough estimate of the
        energies involved.
        Masses quoted for O7.5I stars are on the order of 60 \msunyr\
        implying a main sequence type of O5 or earlier. Again, the
        type O5 was chosen to give a rough estimate of the
        energies involved.
        Their stellar wind parameters were calculated from the
        empirical relations of \citet{how89}. Their main
        sequence lifetime is from \citet{chiosi78}.}
\tablenotetext{[a]}{\citet{van01}.}
\tablenotetext{[b]}{\citet{WR5_vinf}. Values found in
  the literature range from 1600 \kms\ \citep{KH95} to
2800 \kms\ \citep{Torres86}.}
\tablenotetext{[c]}{\citet{KH95}.}
\tablenotetext{[d]}{\citet{HD17603type}.}
\tablenotetext{[e]}{\citet{HD17603_winds}, based on comparison with
        other stars.}
\tablenotetext{[f]}{\citet{HD17603_winds}, based on empirical relations
        by Garmany \& Conti (1984).}
\end{deluxetable}

\clearpage

\begin{deluxetable}{l l l}
\tablewidth{0pt}
\tablecaption{Physical parameters of the \hi\ tails
\label{tbl:tail_param}}
\tablehead{\colhead{Parameter} & \colhead{Notation} & \colhead{Value}}
\startdata
Average \hi\ column density in \hig & ${\cal N}$ & $1.2 \times 10^{20}$ cm$^{-2}$ \\
Total mass of \hig & $M$ & $1700$ \msun \\
Flow time scale for \hig & $\tau$ & $4.7 \times 10^6\ \frac{1}{\tan(i)}$ yr \\
Linear momentum in \hig & $P$ & $4.2 \times 10^{42}\ \frac{1}{\cos(i)}$ \gcms \\
Kinetic energy in \hig & $E$ & $2.7 \times 10^{48}\ \frac{1}{\cos^2(i)}$ erg \\Minimum total kinetic energy from driving sources & $E_{*, min}$ & $1.4 \times
10^{51}\ \frac{1}{\epsilon_E}$ erg \\
Minimum total momentum from driving sources & $P_{*, min}$ & $1.4
\times 10^{45}\ \frac{1}{\epsilon_P}$ g cm s$^{-1}$ \\
Minimum rate of kinetic energy delivery & $\dot{E}_{*, min}$ &
$7.2 \times 10^{36}\ \frac{1}{\epsilon_E}$ erg s$^{-1}$ \\
Minimum rate of momentum delivery & $\dot{P}_{*, min}$ & $9.6
\times 10^{30}\ \frac{1}{\epsilon_P}$ g cm $s^{-2}$ \\
\enddata
\end{deluxetable}

\clearpage

\figcaption[]{Mean {\hi} brightness temperature image of {\hig}, averaged
between LSR radial velocities $-53.4$ and $-66.6$~{\kms}.
\label{fig:mean_Tb_HI}}

\figcaption[]{The velocity gradient along {\hig} is highlighted by the
transition from red ({\hi} emission at $-57.1$~{\kms}), to green
($-60.4$~{\kms}), to blue ($-63.7$~{\kms}) as one moves along along the tails.
The width of the velocity bins is $3.3$~{\kms}.
\label{fig:vel_gradient}}

\figcaption[]{Eight {\hi} channel maps of {\higfull} showing filamentary
substructures. The positions of WR~5 and HD~17603 are indicated. Each velocity
channels spans 0.82~{\kms}. Greyscale runs from 0~{\kelvin} (white) to
75~{\kelvin} (white). 
\label{fig:channel_maps}}

\figcaption[]{Contours of {\coi} at $-46.81$~{\kms} at 0.75, 1.25, and
2.0~{\kelvin} superimposed on a greyscale of the {\hi} at $-58.35$~{\kms}. The
northern narrow {\hi} filament appears to emanate from a faint CO cloudlet
within a gap in a chain of brighter CO clouds. The thicker southern part of the
{\hi} tails appear to emanate from {\cog}.
\label{fig:co_clouds}}

\figcaption[]{Dependance of $P_*$ (------), $E_*$ ($\cdot\, \cdot\, \cdot\,
\cdot $), $\dot{P}_*$ ($-\, -\, -\,  - $), and $\dot{E}_*$ ($- \cdot -
\cdot -$) on the inclination angle $i$. The first two parameters are scaled by
$\frac{\epsilon_{_X}}{X_{_0}}$ and the last two by $\frac{\tau_{_0}
\epsilon_{_X}}{X_0}$, where $X$ is $P$ or $E$.
\label{fig:dep_on_i}}

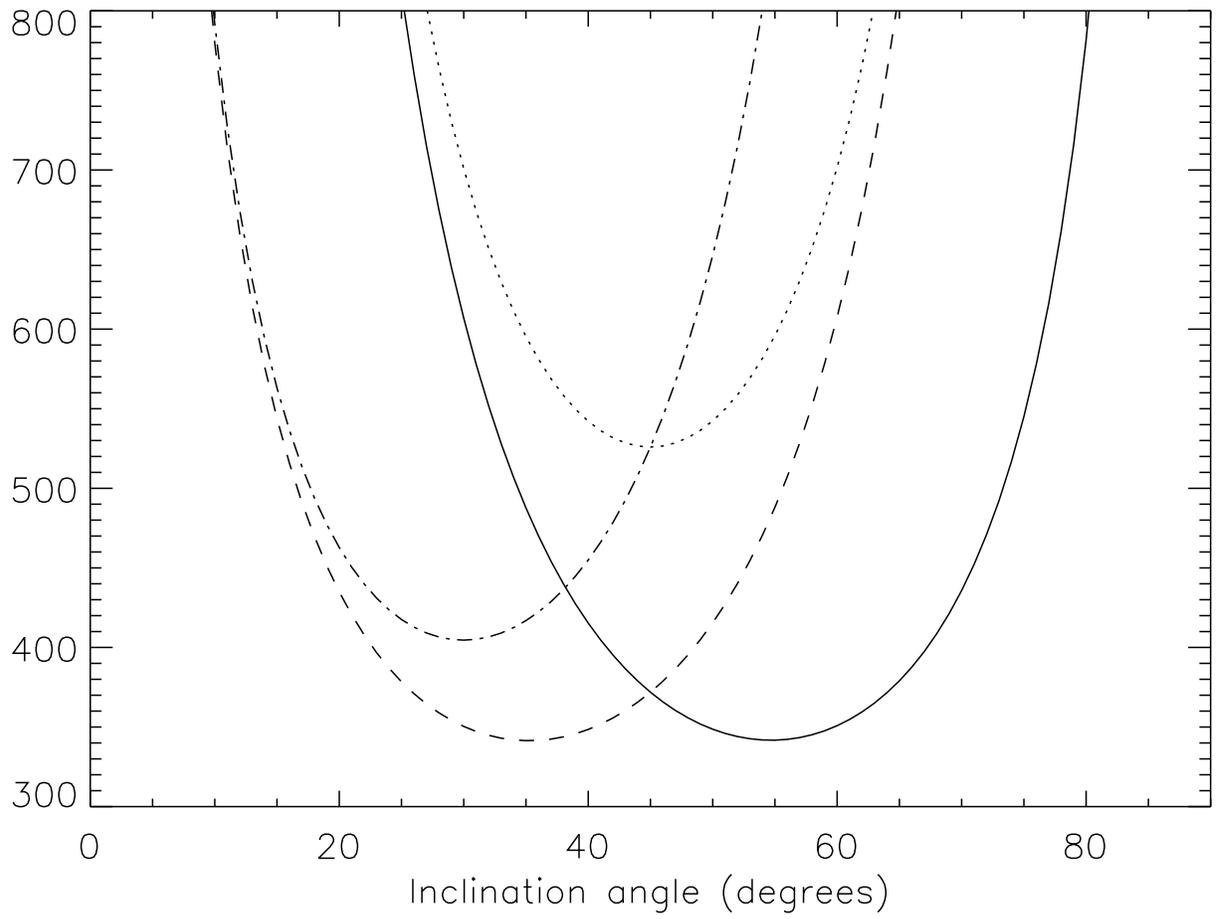
\end{document}